\documentclass[pss]{w-art}

\usepackage{url}
\usepackage{times}
\usepackage{w-thm}
\usepackage[dvips]{color,epsfig,graphicx}
\usepackage{amsmath,amssymb,amsfonts}
\usepackage{citesort}
\usepackage{sidecap}

\theoremstyle{plain}

\newcommand{\ZZ}{\mathbb{Z}}

\newcommand{\NN}{\mathbb{N}}

\newcommand{\spa}{{\rm span}}

\newcommand{\la}{\langle}
\newcommand{\ra}{\rangle}

\renewcommand{\phi}{\varphi}

\newcommand{\intd}{\mathrm{d}} %upright d in integrations
\newcommand{\vect}[1]{\boldsymbol{#1}}
\newcommand{\ml}{\ell} 	%magnetic length
\newcommand{\half}{\frac{1}{2}}
\renewcommand{\i}{\mathrm{i}}		%imaginary i as an upright letter
\newcommand{\ex}{\vect{e}_\mathrm{x}}	 %unit vector in the x-direction
\begin{document}

%%%%%%%%%%%%%%%%%%%%%%%%%%%%%%% Title page 
\pagespan{1}{}
%\kywords{..}
% 1. Quantum Hall Effects. Theory and Modeling
% 2. Fermion Systems and electron gas (part of quantum statistical physics)

\title[Laughlin's function on a cylinder -- plasma analogy and quantum polymer]
	{Laughlin's function on a cylinder: plasma analogy and representation as a quantum polymer}

\author[S. Jansen]{Sabine Jansen\footnote{Corresponding
     author: e-mail: {\sf jansen@math.tu-berlin.de}}\inst{1}}
 	\address[\inst{1}]{Institut f\"ur Mathematik, Technische Universit\"at Berlin, Stra\ss e 
	des 17. Juni 136, 10623 Berlin, Germany}
\author[E.\thinspace H. Lieb]
	{Elliott H. Lieb \footnote{e-mail: {\sf lieb@princeton.edu}}\inst{2}}
	\address[\inst{2}]{Princeton University, Jadwin Hall, P.O. Box 708, Princeton, 
	NJ 08542-0708, USA}
\author[R. Seiler]{Ruedi Seiler \footnote{e-mail: {\sf seiler@math.tu-berlin.de}}\inst{1}}

\begin{abstract} 
	We investigate Laughlin's fractional quantum Hall effect
	 wave function in the cylinder geometry of Laughlin's 
	integer quantum Hall effect argument, at filling factor $1/3$. We show that the plasma analogy leads 
	to a periodic density, and that 
	the wave function admits a representation as a ``quantum polymer'', 
	reminiscent of the quantum dimer model by Rokhsar and Kivelson. We 
	explain how the representation can be exploited to compute the normalization
	and one-particle density in the limit of infinitely many particles. 
\end{abstract}

\maketitle

%%%%%%%%%%%%%%%%%%%%%%%%%%%%%%%%%%%%% main matter %%%%%%%%%%%%%%%%%
\section{Introduction}

Laughlin's wave function \cite{l83} is widely accepted as a good description of fractional Hall effect
 ground states at simple filling fractions. It was initially proposed for a disk geometry, but later 
adapted to various geometries \cite{thousurfsci,hr85,hr85sphere}. In this article, we study 
Laughlin's function on a cylinder \cite{thousurfsci}.
The choice of geometry is motivated 
by its use in Laughlin's argument \cite{l81} for the integer quantum Hall effect (IQHE) 
and the suggestion that a ground state degeneracy is required in order to reconcile Laughlin's 
IQHE argument with fractional charge transport \cite{tw84}.

Ground state degeneracy, in turn, is closely related to 
broken translational symmetries and multiple Aharonov-Bohm periods (see, e.g., \cite{t89,thoulessgefen}, and 
Sec.~\ref{sec:aha}). In an earlier paper \cite{jls}, we have shown that Laughlin's state 
on thin cylinders, at filling factor $1/3$, is periodic, with period three times that 
of the filled Landau level. In this article, we present considerations complementary 
to \cite{jls}. 

A first aspect pertains to the plasma analogy, which relates the modulus squared of Laughlin's 
wave function  to the Boltzmann weight of a classical two-dimensional one-component plasma, or jellium. 
It was used in \cite{l83} to justify that Laughlin's wave function in the disk, at filling 
fraction $1/3$, has a homogeneous density.  In Sec.~\ref{sec:plasma}, we explain that the 
same analogy, as soon as it is applied to the cylinder, leads to a periodic density. This uses
results on jellium on cylinders (``jellium tubes'') \cite{swk04}.

A second aspect focuses on a detailed analysis of the structure of Laughlin's wave 
function derived from combinatorial properties of powers of Vandermonde determinants 
\cite{d93,fgil94}. We show that Laughlin's wave function admits a representation
 that we call a ``quantum polymer'' because of its resemblance with the 
quantum dimer model \cite{rk88} (Sec.~\ref{sec:qupoly}). 
It turns out that this representation is very useful to access the normalization and 
the one-particle density: In Secs.~\ref{sec:norm} and~\ref{sec:density}, we review the results on normalization 
and the density's periodicity obtained in \cite{jls} and propose a computational scheme, based 
on the polymer representation, 
for the normalization and the density in  the limit of infinitely many particles.
Roughly, 
we obtain an expansion in powers of $\exp(-\gamma^2)$, where $\gamma$ is the ratio 
of the magnetic length and the cylinder radius. 
The zeroth order in the expansion is the Tao-Thouless state \cite{tt83}, which is $3$-periodic 
and corresponds to the limit of infinitely thin cylinders. 
This is closely related to results by Rezayi and Haldane \cite{rh94}. However, periodicity 
does not stop at order zero. In fact, we obtain a sequence of approximate densities, each of which 
is periodic with smallest period $3$. 
% Hence, the quantum polymer representation provides 
% a practical way to access the density as well as a way to understand the periodicity.

Hence, in the FQHE context, the quantum polymer representation is useful in two ways. First, 
it shows that a $3\gamma$-periodicity is indeed built-in into Laughlin's wave function on a cylinder, 
as is expected \cite{tw84} from Laughlin's IQHE argument. 
% Let us mention that the periodicity 
%  has been noticed much earlier for toroidal geometries \cite{hr85,h85}, 
% but was controversial on a cylinder \cite{taothouless}. 
Second, the quantum polymer formalism 
gives a practical way to access the one-particle density of Laughlin's function. Our method can 
actually be extended to handle correlation functions. This is interesting in view of the widespread 
use of Laughlin's function in theoretical investigations of FQHE ground states.

\section{Symmetry breaking, ground state degeneracy and multiple Aharonov-Bohm periods}
\label{sec:aha}

Before we turn to the analysis of Laughlin's wave function,
 let us briefly recall 
 the relationship between symmetry breaking, 
ground state degeneracy and multiple Aharonov-Bohm periods in the cylinder 
geometry of Laughlin's IQHE article \cite{l81}. 
\begin{vchfigure}[here] 
	\begin{center}
		\resizebox{8cm}{!}{\begin{picture}(0,0)%
\includegraphics{fig1.pstex}%
\end{picture}%
\setlength{\unitlength}{4144sp}%
\begingroup\makeatletter\ifx\SetFigFont\undefined%
\gdef\SetFigFont#1#2#3#4#5{%
  \reset@font\fontsize{#1}{#2pt}%
  \fontfamily{#3}\fontseries{#4}\fontshape{#5}%
  \selectfont}%
\fi\endgroup%
\begin{picture}(8155,3121)(136,-3260)
\put(3691,-3211){\makebox(0,0)[lb]{\smash{{\SetFigFont{12}{14.4}{\rmdefault}{\mddefault}{\updefault}{\color[rgb]{0,0,0}$(3N-3)\gamma$}%
}}}}
\put(4577,-921){\makebox(0,0)[lb]{\smash{{\SetFigFont{14}{16.8}{\rmdefault}{\mddefault}{\updefault}{\color[rgb]{0,0,0}$y$}%
}}}}
\put(7617,-1171){\makebox(0,0)[lb]{\smash{{\SetFigFont{14}{16.8}{\rmdefault}{\mddefault}{\updefault}{\color[rgb]{0,0,0}$\phi$}%
}}}}
\put(4591,-421){\makebox(0,0)[lb]{\smash{{\SetFigFont{12}{14.4}{\rmdefault}{\mddefault}{\updefault}{\color[rgb]{0,0,0}$\vect{B}$}%
}}}}
\put(2476,-2851){\makebox(0,0)[lb]{\smash{{\SetFigFont{12}{14.4}{\rmdefault}{\mddefault}{\updefault}{\color[rgb]{0,0,0}$\gamma$}%
}}}}
\put(3421,-1276){\makebox(0,0)[lb]{\smash{{\SetFigFont{14}{16.8}{\rmdefault}{\mddefault}{\updefault}{\color[rgb]{0,0,0}$x$}%
}}}}
\put(7968,-2401){\makebox(0,0)[lb]{\smash{{\SetFigFont{14}{16.8}{\rmdefault}{\mddefault}{\updefault}{\color[rgb]{0,0,0}$x$}%
}}}}
\put(4617,-2501){\makebox(0,0)[lb]{\smash{{\SetFigFont{14}{16.8}{\rmdefault}{\mddefault}{\updefault}{\color[rgb]{0,0,0}$...$}%
}}}}
\put(1306,-2446){\makebox(0,0)[lb]{\smash{{\SetFigFont{14}{16.8}{\rmdefault}{\mddefault}{\updefault}{\color[rgb]{0,0,0}$\frac{\phi}{2\pi} \gamma$}%
}}}}
\put(1306,-781){\makebox(0,0)[lb]{\smash{{\SetFigFont{14}{16.8}{\rmdefault}{\mddefault}{\updefault}{\color[rgb]{0,0,0}$R=1/\gamma$}%
}}}}
\put(136,-1816){\makebox(0,0)[lb]{\smash{{\SetFigFont{12}{14.4}{\rmdefault}{\mddefault}{\updefault}{\color[rgb]{0,0,0}$|\psi_k(z;\phi)|^2$}%
}}}}
\end{picture}%
}
	\end{center}
	\vchcaption{\label{fig:chargetransport} \label{fig:LLL} 
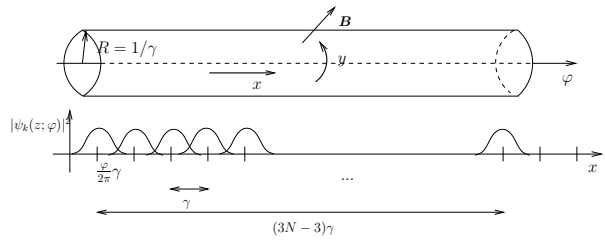
		Cylinder geometry and lowest Landau level basis functions. The basis functions 
		are Gaussians with centers spaced apart by $\gamma$. 
		Laughlin's state occupies a cylinder of length roughly $(3N-3)\gamma$. 
	}
\end{vchfigure}

Consider electrons moving on a cylinder of radius $R$ in 
a perpendicular magnetic field of strength $B$ (see Fig.~\ref{fig:LLL}). 
The cylinder is threaded 
by an additional flux $\phi$, which does not change the magnetic field $B$ but 
changes the vector potential. Each particle position is described 
by its coordinate $x$ along the cylinder axis, and by an angular 
coordinate $y\in [0,2\pi R[$. It is convenient to choose units such that 
Planck's constant $\hbar$, the electron mass $m_\mathrm{e}$, the elementary 
charge $e$ and the field magnitude $B$ all take the value $1$. In those units, 
the magnetic length $\ml = (\hbar /eB)^{1/2}$ equals $1$ and the flux 
quantum is $\phi_0 = h/e = 2\pi$. A crucial role is played 
by the dimensionless parameter $\gamma = \ml /R$ which in our units
becomes the inverse 
of the radius, $\gamma = 1/R$. 
In the absence of a background potential, the one-particle Hamiltonian for the 
infinite cylinder is
\begin{equation} \label{eq:ham}
 	H(\phi) = \frac{1}{2}\bigl[ -\partial_x ^2 + (-\i \partial_y 
	- x + \frac{\phi\gamma}{2\pi})^2\bigr]
\end{equation}
with periodic boundary conditions in the $y$-direction. 

\begin{SCfigure}
	\resizebox{6cm}{!}{\input{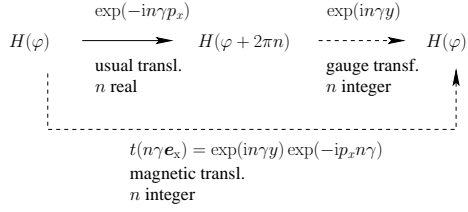}}
	\caption{Gauge periodicity and discrete translational invariance for the Landau Hamiltonian~(\ref{eq:ham}) on 
		a cylinder:
		a usual translation by $n\gamma$ changes the flux from $\phi$ to $\phi + 2\pi n$ 
		(full arrow). When $n$ is an integer (dotted arrows), the change in flux can be gauged away,
		 leading to the invariance of the Hamiltonian with respect to the 
		\emph{magnetic} translation.}
	\label{fig:gaugeinv}
\end{SCfigure}

The family of Hamiltonians $H(\phi)$ has two intimately related features,
\emph{gauge periodicity} and a \emph{discrete} translational invariance, 
see Fig.~\ref{fig:gaugeinv}. 
Both aspects become explicit in the lowest Landau level, i.e., 
the ground state of the Hamiltonian~(\ref{eq:ham}). It has a complete orthonormal set 
\begin{equation*}
 	\psi_k(z;\phi) = \frac{1}{\sqrt{2\pi \gamma^{-1}\sqrt{\pi}}} 
		\exp(\i k\gamma y) \exp\{ -\frac{1}{2}[ x-(k+\phi/2\pi)\gamma]^2\},\ z=x+\i y,\ k\in \ZZ.
\end{equation*}
Notice that $|\psi_k(z;\phi)|^2$, $k\in \ZZ$, are Gaussians of $x$ with centers spaced 
apart by $\gamma$. Consequently, the density in the filled Landau level is a sum of equally weighted Gaussians. 
Its Fourier series can be computed using 
Poisson's summation formula,
\begin{equation*}
 	\rho_\mathrm{filled\ LLL}(z;\phi) = \sum_{k=-\infty}^\infty |\psi_k(z;\phi)|^2 = 
		\frac{1}{2\pi}\Bigl\{1+ 2 \sum_{k=1}^\infty \exp(-\frac{\pi^2 k^2}{\gamma^2})
		\cos[k(\frac{2 \pi  x}{\gamma} - \phi)]\Bigr\}.
\end{equation*}
This formula shows that the filled Landau level density is always periodic 
with minimal period $\gamma$, although the amplitude of oscillations becomes very small 
when $\gamma$ is small.\footnote{The amplitude is, roughly, $\exp(-\pi^2 /\gamma^2)= 
\exp(-\pi^2 R^2)$. A similar small $\gamma$ behavior 
for the amplitude of oscillations of the $3\gamma$-periodic 
density at filling factor $1/3$ has been found numerically in \cite{leeleinaastorus}.}
Changing $\phi$ amounts to an overall shift of the density. 
After a change of $\phi$ by $2\pi$, the density returns 
to itself: the translational period $\gamma$ is associated with the simple Aharonov-Bohm period 
$\Delta \phi = 2\pi $. 

Now suppose that a ground state has translational period $3\gamma$ 
instead of $\gamma$. Then it takes $3$ flux units before the state returns to itself. Because of the 
system's gauge periodicity, the states obtained after addition of one or two 
flux units are ground states as well. Hence, $3\gamma$-periodicity 
leads not only to a threefold Aharonov Bohm period but also to threefold degeneracy of 
ground states.

\section{Plasma analogy} \label{sec:plasma}

Laughlin invoked the plasma analogy to show that the density of his wave 
function, in a disk (``Laughlin droplet''), is constant. In this section, we show 
that the same analogy, when it is invoked in a cylinder geometry, leads to a one-particle 
density that is periodic with period $3\gamma$. This is based on earlier observations 
that jellium tubes should have a periodic density \cite{agl01,swk04}. For plasma parameter 
$\Gamma =2$, this was already shown in \cite{cfs83}.

Let us start by a brief description of the system we are interested in. We consider 
$N$ particles of charge $q$ moving on a strip $[-L/2,L/2]\times [0,2\pi R]$
and interacting with a neutralizing background of uniform charge density $-nq$. 
We work with complex coordinates $z = x+\i y$ and choose a semi-periodic logarithmic ``Coulomb'' potential
$V(z):= -\ln \left| 2\sinh [z/(2R)] \right|$. The potential energy is 
\begin{multline*}
\Phi (z_1,...,z_N)= q^2 \sum_{1\leq j<k\leq N} V(z_j-z_k) -nq^2 \sum_{j=1}^N \int_{-L/2}^{L/2}
		 \int_0^{2\pi R} V(z_j -z) \intd x \intd y \\
	+\half n^2q^2 \int_{[-L/2,L/2]^2}\int_{[0,2\pi R]^2} V(z-z') \intd x \intd y 
	\intd x' \intd y'.
\end{multline*}
For the special choice that the plasma parameter $\Gamma:=\beta q^2$ equals twice the exponent 
in Laughlin's function, here, $\Gamma = 2\cdot 3$,  
the Boltzmann weight $\exp(-\beta \Phi)$ becomes proportional to Laughlin's function.
More precisely, let us choose units 
such that the background density is $n=(3\cdot 2\pi)^{-1}$, and suppose that $\Gamma =6$.
Then 
\begin{equation} \label{eq:analogy}
	\exp\bigl(-\beta \Phi(z_1,...,z_N)\bigr) \propto 
		\Bigl|\Psi_N\bigl( \{ z_j - \frac{3(N-1)}{2}\gamma\} \bigr)\Bigr|^2.
\end{equation}
Here, $\Psi_N$ is Laughlin's cylinder function with magnetic length $\ml=1$, see 
Eq.~(\ref{eq:laufunct}).
Therefore the plasma system and Laughlin's function will have the same one-particle density. 

For a plasma on a plane, at coupling $\Gamma =6$, we expect a homogeneous density. 
On the other hand, a cylinder is more like a quasi one-dimensional structure, especially when 
we consider long cylinders \cite{agl01}. Notice also that the $y$ average of the 
potential $V(z)$ equals $-|x|/(2R)$, i.e., a one-dimensional Coulomb potential. 
As was observed in \cite{swk04}, the comparison with one-dimensional jellium leads 
us to the expectation that the density is periodic in the direction of the cylinder axis 
with period $L/N$.  This is because one-dimensional jellium 
 always displays symmetry breaking \cite{k74,bl75}. 
The interpretation is ``Wigner crystallization'', the electrons tend to minimize their repulsive interaction 
by adopting positions on a lattice with spacing $L/N$. 

Now, the crucial point is that the period $L/N$ expected for jellium on a cylinder translates 
into $3\gamma$-periodicity for Laughlin's function on a cylinder. Indeed, the cylinder's length 
is 
$$L=\frac{N}{2\pi R n} = \frac{N}{2\pi \gamma^{-1}\,(3\cdot 2\pi)^{-1} }= 3\gamma N.$$
Thus because of the plasma analogy, we expect that the density in Laughlin's function is periodic with 
period $3\gamma$, i.e., three times the period of the filled Landau level. 
The periodicity is confirmed by 
numerical results on jellium tubes \cite{swk04}, from which it appears that the amplitude of oscillations 
is small when the cylinder radius gets large. This fits nicely into the picture that the 
cylinder interpolates between a line and a plane. The important point is that 
the cylinder retains the periodicity of the one-dimensional system. 

\section{Laughlin's wave function as a quantum polymer}\label{sec:qupoly}

In the previous section, we have used the plasma analogy to argue
 that a $3\gamma$-periodicity in the axial direction is very natural. In
 this section, we  take a different approach and investigate 
directly the structure of the wave function. This leads us 
to a recursive pattern in the sequence of Laughlin functions $(\Psi_N)_{N\in \NN}$, formalized with 
a ``product rule'', and to the representation of Laughlin's function as 
a ``quantum polymer''.

We use the units and notation of Sec.~\ref{sec:aha}, with the flux parameter 
$\phi$ set to $0$. 
Laughlin's function adapted to the cylinder geometry \cite{thousurfsci} 
at filling factor $1/3$ is
\begin{equation}\label{eq:laufunct}
	\Psi_N(z_1,...,z_N)= 
	\kappa_N \prod_{1\leq j<k\leq N} \bigl(\exp(\gamma z_k)-\exp(\gamma 
		z_j)\bigr)^3 \exp(-\sum_{j=1}^N x_j^2/2).
\end{equation}
It will be convenient to fix the multiplicative constant $\kappa_N$ as 
\begin{equation} \label{eq:multconst}
	\kappa_N = \frac{1}{\sqrt{N!}} \frac{1}{(2\pi \gamma^{-1} \sqrt{\pi})^{N/2}} 
		\exp(-\frac{9}{2} \gamma ^2 
		\sum_{j=0}^{N-1} j^2 ).
\end{equation}

\subsection{Product rule}
The wave function $\Psi_N$ is  
a polynomial of $\exp(\gamma z_1), ..., \exp(\gamma z_N)$ times a Gaussian weight. 
Expanding the polynomial, we find an expression of $\Psi_N$ in terms 
of lowest Landau level basis functions $\psi_k(z) = \psi_k(z;0)$ given 
in Sec.~\ref{sec:aha}. For up to three particles, this gives
\begin{equation}\label{expex}
	\begin{aligned}
		\Psi_1 & = \psi_0,\\
		\Psi_2 & = \psi_0 \wedge \psi_3 - 3 e^{-2\gamma^2} \psi_1\wedge \psi_2\\
		\Psi_3 & = \psi_0\wedge \psi_3 \wedge \psi_6 - 3 e^{-2 \gamma^2}\psi_0\wedge \psi_4\wedge \psi_5
	 	 - 3 e^{-2 \gamma^2} \psi_1\wedge \psi_2\wedge \psi_6 \\
			&\quad - 6 e^{-5\gamma^2} \psi_1\wedge \psi_3 \wedge \psi_5 
			+ 15 e^{-8\gamma^2} \psi_2\wedge \psi_3 \wedge \psi_4.
	\end{aligned}
\end{equation}
Notice that the coefficient $-3 \exp(-2\gamma^2)$ in the expansion of $\Psi_2$ reappears 
in the expansion of $\Psi_3$. Similarly, the coefficients in the expansion of $\Psi_3$ 
will show up in the expansion of $\Psi_4$. The aim of this section is to explain and formalize 
this phenomenon.

In general, one can write down an expansion 
\begin{equation} \label{eq:generalex}
	\Psi_N  = \sum_{0\leq m_1<...<m_N\leq 3N-3} a_N(m_1,...,m_N) \psi_{m_1}\wedge ...\wedge \psi_{m_N}.
\end{equation}
As can be seen in Eqs.~(\ref{expex}), 
the expansion coefficients $a_N(\vect{m})$ are always made up of two parts, 
an integer and a power of $\exp(-\gamma^2)$.  
The integer comes from the expansion of 
the third power of a Vandermonde determinant. The power of $\exp(-\gamma^2)$  
accounts for the multiplicative constant $\kappa_N$ 
and for the normalization of $\exp(\gamma k z) \exp(-x^2/2)$.
More precisely,
\begin{equation*}
	a_N(m_1,...,m_N) = b_N(m_1,...,m_N) \exp\bigl(\half \gamma^2\sum_{j=1}^N (m_j^2 - 9(j-1)^2)\bigr)
\end{equation*}
with suitable integers $b_N(\vect{m})$. 
 The coefficients 
$a_N(\vect{m})$ are specific to the cylinder geometry and have been analyzed by Rezayi and Haldane~\cite{rh94}, while the integers $b_N(\vect{m})$ play a role as well for the usual ``Laughlin droplet'' geometry and have 
inspired a certain amount of combinatorial research~\cite{d93,fgil94,stw94,ktw01}.

It is practical to associate with each finite sequence 
$\vect{m}$ a sequence of $1$'s and $0$'s characterizing occupancy or non-occupancy of 
lattice sites $\{-1,0,1,...,3N-2\}$. Each site $k$ stands for the orbital $\psi_k$.
The sites $-1$ and $3N-2$ will never be occupied, but it is advantageous to include 
them nevertheless.  For $N=1$, there is only one configuration, $010$, corresponding 
to $m_1 =0$. For $N=2$, the two wedge products give rise to 
\begin{equation*}
 (0,3)\ \hookrightarrow\ (010\mid 010),\quad (1,2)\ \hookrightarrow (001100).
\end{equation*}
Observe that $(0,3)$ corresponds to a concatenation of two identical 
one-particle blocks $010$, as highlighted by the vertical bar. Similarly, for $N=3$, 
one- and two-particle blocks show up: 
\begin{equation}\label{eq:block3}
  \begin{alignedat}{2}
	(0,3,6)\ &\hookrightarrow\ (010\mid 010\mid 010),&\quad 
	(1,2,6)\ &\hookrightarrow\ (001100 \mid 010),\\
	(2,3,4)\ &\hookrightarrow\ (000111000),&\quad
	(0,4,5)\ &\hookrightarrow\ (010\mid 001100),
  \end{alignedat}
\end{equation}
etc. Now, we assign to each block the amplitude of the corresponding $\vect{m}$.
Thus
\begin{equation*}
	a_1(010) = a_1(m_1=0) =1,\quad a_2(001100)= a_2(1,2)= - 3 e^{-2\gamma^2}.
\end{equation*}
The crucial observation is the following \emph{product rule}:
\begin{quote}
	\emph{Concatenation of blocks results in the multiplication of amplitudes.}
\end{quote}
For example, 
\begin{equation} \label{eq:prodrulex}
   \begin{aligned}
	a_3(010\mid 010 \mid 010)& = a_3(0,3,6)=1 = a_1(010)a_1(010)a_1(010),\\
	a_3(010\mid 001100) &= a_3(0,4,5)= -3e^{-2\gamma^2} = a_1(010) a_2(001100).
   \end{aligned}
\end{equation}
A rigorous proof of the product rule for $a_N(\vect{m})$ is given in \cite{jls}, 
the product rule for the integer part $b_N(\vect{m})$ has been shown earlier \cite{fgil94}. 

There is a simple characterization of the block decomposition of a sequence of occupation 
numbers corresponding to a vector $\vect{m}$ in terms of the 
differences
\begin{equation} \label{eq:partialsum}
	\nu_k:=\sum_{j=1}^k m_j- \sum_{j=1}^k 3(j-1),\quad k\in \{1,...,N\}.
\end{equation}
When $\vect{m}$ has a non-vanishing amplitude $a_N(\vect{m})$, the $\nu_k$'s 
are non-negative and $\nu_N=0$ \cite{fgil94}. 
The block decomposition is determined by the additional zeros of $(\nu_k)$: 
each bar after $3k$ sites corresponds to a vanishing difference, i.e., $\nu_k=0$. 
This is the characterization used in \cite{fgil94,jls}.

A different characterization is in terms of  squeezing operations  \cite{rh94}. 
Rezayi and Haldane \cite{rh94} observed that every $\vect{m}$ with non vanishing amplitude 
$a_N(\vect{m})$ can be obtained from $(0,3,6,...,3(N-1))$ by a series of elementary operations 
$(a_1,...,a_N) \mapsto (a_1,...,a_i+1,...,a_j-1,...,a_N)$, $i<j$, or ``squeezings''. If for 
some $k$ we only consider sqeezings that respect a boundary at $k$, i.e., never squeeze pairs 
$(i,j)$ with $i<k$ and $k<j$, then the resulting $\vect{m}$ will split into two 
blocks of $k$ and $N-k$ particles. \\

It is important to notice that the product rule contains, implicitly, a 
periodicity statement: 
 the amplitude assigned to a block does not depend on its position 
in the sequence of occupation numbers. In Eq.~(\ref{eq:prodrulex}), 
$010$ has amplitude $1$ regardless whether it refers to a particle sitting 
in the site $0$, $3$, or $6$.
This periodicity is at the origin of $3\gamma$-periodicity 
in Laughlin's state. \\

\begin{remark}[Reversal invariance] \label{rem:revinv}
In addition to the product rule, 
there is another property of the coefficients $b_N(\vect{m})$, shown in~\cite{d93,fgil94}, 
that transfers to 
the amplitudes $a_N(\vect{m})$: replacing $\vect{m}$ with the reversed sequence 
$(3N-3-m_{N-j})_j$ leaves the amplitude unchanged. This means that Laughlin's cylinder 
function is unchanged by overall rotations by $180^\circ$ around the middle of the cylinder, 
$(3N-3)\gamma/2$.
\end{remark}

\subsection{Quantum polymer}

Now we translate
the product rule for the amplitudes $a_N(\vect{m})$ 
into a representation formula for the function $\Psi_N$. 
As a first step, notice that the way a configuration splits into blocks 
defines a partition of the discrete volume $\Lambda_N = \{-1,0,...,3N-2\}$ into discrete intervals, 
or ``rods'', $X_j$. For example,
\begin{equation*}
    \vect{m}=(0,3) \triangleq (010\mid 010)\ \hookrightarrow\ \Lambda_2=\{-1,0,1\} \cup \{2,3,4\} = X_1 \cup X_2.
\end{equation*}
In the expansion~(\ref{eq:generalex}), we can group together 
$\vect{m}$'s that belong to the same partition. Using the product rule, 
we obtain a representation of $\Psi_N$ as a sum over partitions. For $N=3$ particles,
 this gives 
\begin{equation}\label{eq:expn3}
\begin{aligned}
	\Psi_3  &= u_{\{-1,0,1\}}\wedge u_{\{2,3,4\}}\wedge u_{\{5,6,7\}} \\
		&\quad + u_{\{-1,0,1\}}\wedge u_{\{2,...,7\}} + u_{\{-1,...,4\}}\wedge u_{\{5,6,7\}}
			+ u_{\{-1,...,7\}}
\end{aligned}
\end{equation}
with the functions $u_X$ from Table~\ref{tab:ux}.
\begin{vchtable}
	$\begin{array}{ccll}
		\hline \hline
		N(X) & X & u_X & \alpha_{N(X)}=||u_X||^2 \\
		\hline 
		1 & \{-1,0,1\}& \psi_0 &1\\
		 & \{2,3,4\} & \psi_3 &1 \\
		 & \{5,6,7\} & \psi_6 &1 \\
		\hline
		2 & \{-1,...,4\} & -3 e^{-2\gamma^2} \psi_1\wedge \psi_2 & 9e^{-4\gamma^2}\\
		 & \{2,...,7\} & -3 e^{-2\gamma^2}\psi_4 \wedge \psi_5 & 9e^{-4\gamma^2} \\
		\hline
		3 & \{-1,...,7\} & - 6 e^{-5\gamma^2} \psi_1\wedge \psi_3 \wedge \psi_5 
				& 36 e^{-10\gamma^2} + 225 e^{-16\gamma^2} \\
		 & &\quad	+ 15 e^{-8\gamma^2} \psi_2\wedge \psi_3 \wedge \psi_4 &\\
		\hline \hline
	  \end{array}$
	\vchcaption{ \label{tab:ux}
		Polymer functions $u_X$ contributing to the representation~(\ref{eq:expn3}) of Laughlin's 
		wave functions for $N=3$ particles. 
	}	
\end{vchtable}
More generally, with each rod $X$ of cardinality (or ``length'') a multiple of $3$, 
we associate a fermionic wave function $u_X$ of $|X|/3$ complex variables. 
$u_X$ describes a cloud of electrons of density $1/3$, localized in the discrete volume $X$: 
\begin{equation*}
	u_X \in \wedge^{N(X)} \spa \{ \psi_k \mid k\in X\}, \quad N(X) = |X|/3.
\end{equation*}	
Shifting a rod amounts to a magnetic translation of the function: 
\begin{equation}\label{eq:covar}
	\qquad u_{3j+X} = t(j 3\gamma\,\ex )^{\otimes N(X)}u_X,\quad j\in \ZZ.
\end{equation}
This translational covariance is a consequence of the periodicity 
in the amplitudes, mentioned in the previous subsection. 
With the functions $u_X$, Eq.~(\ref{eq:expn3}) is generalized as follows:
\begin{quote} \it 
 	Laughlin's wave function $\Psi_N$ at filling factor $1/3$ is a 
	sum over ordered partitions of the discrete volume $\{-1,0,...,3N-2\}$ into rods $X_j$ with lengths 
	that are multiples of $3$:
	\begin{equation} \label{eq:part}
		\Psi_N = \sum_{(X_1,...,X_D)} u_{X_1}\wedge ... \wedge u_{X_D}.
	\end{equation}
\end{quote}
A partition is \emph{ordered} when the rods $X_j$ are labelled from left to right.
In Eq.~(\ref{eq:part}), different partitions give actually rise to orthogonal contributions. 

The representation~(\ref{eq:part}) is reminiscent of the quantum dimer (QD) model proposed 
by Rokhsar and Kivelson~\cite{rk88} as an idealization to resonating valence bond states.
QD wave functions are given as sums over partitions 
of a lattice into dimers (i.e., subsets of two neighbor elements). 
Each dimer represents  two spin $1/2$ particles that form a singlet. 
Motivated by the quantum dimer model, we will say that Laughlin's function can be expressed 
as a (discrete) \emph{quantum polymer}. The associated 
polymer system consists of $3n$-mers and has a built-in periodicity, see Eq.~(\ref{eq:covar}).

In the next sections, we will exploit the quantum polymer representation to obtain 
information on the normalization and the density of Laughlin's function. 

\section{Normalization}\label{sec:norm}

For computational purpose it is useful to have information on the normalization constant 
$C_N = ||\Psi_N||^2$ 
of the wave function. In \cite{fgil94},
 the plasma analogy is used to derive results on the asymptotics of the normalization 
in a disk geometry. 
Here, a similar approach shows that 
\begin{equation*}
 	\lim_{N\rightarrow\infty} \frac{1}{N}\log C_N = - \log r 
\end{equation*}
for some $r>0$. The limit is closely related to the free energy of the associated 
classical plasma, see Sec.~\ref{sec:plasma}. 
The quantum polymer representation allows us to give more refined results, namely the convergence
\begin{equation*}
 	\lim_{N\rightarrow \infty} r^N C_N = q \geq 0.
\end{equation*}
The quantities $r$ and $q$ are determined by the polymer functions $(u_X)$ in a simple way. 
Because of the translational covariance~(\ref{eq:covar}), the norms $||u_X||$ depend on $N(X)$ only 
and we can define non-negative numbers $(\alpha_n)_{n\geq 1}$ by 
\begin{equation*}
	\alpha_{N(X)} := ||u_X||^2, 
\end{equation*}
see also Table~\ref{tab:ux}. The polymer representation leads to a formula for the normalization,
\begin{equation} \label{eq:normal}
 	C_N = ||\Psi_N ||^2 = \sum_{(X_1,...,X_D)} ||u_{X_1}||^2\cdot ... \cdot ||u_{X_D}||^2 
		= \sum_{\substack{D,n_1,...,n_D:\in \NN \\n_1+...+n_D = N}}
		 \alpha_{n_1}\cdot... \cdot \alpha_{n_D}.
\end{equation}
It follows that $(C_N)$ satisfies a recurrence
 known in stochastics as a discrete 
renewal equation \cite{feller}, with coefficients $(\alpha_n)$ (see Eq.~(\ref{eq:renewal})).
 As a consequence, one of the 
following cases necessarily holds: 
\begin{enumerate}
 	\item $\sum_{n=1}^\infty r^n \alpha_n =1$,\ $\mu:=\sum_{n=1}^\infty nr^n \alpha_n <\infty$,\ 
		$q= \mu^{-1}>0$
	\item $\sum_{n=1}^\infty r^n \alpha_n =1$,\ $\mu:=\sum_{n=1}^\infty nr^n \alpha_n =\infty$,\ 
		$q= 0$
	\item $\sum_{n=1}^\infty r^n \alpha_n <1$,\ $q=0$. 
\end{enumerate}
In the first two cases, $r$ is the unique solution of the equation $\sum_n t^n \alpha_n =1$. In 
the third case, this equation has no solution and $\sum_n t^n \alpha_n$ diverges as soon as $t>r$. 

A probabilistic interpretation can be found in \cite{feller}. We give an interpretation 
in terms of the associated polymer system. Eq.~(\ref{eq:normal}) says $C_N$ is a polymer partition 
function with activities $(\alpha_n)$, see \cite{gruberkunz}. In this context, the rescaling 
of activity in the way $\alpha_n \mapsto r^n \alpha_n$ is very natural, since it leaves 
polymer correlation functions unchanged -- much in the same way as quantum-mechanical expectation 
values $\la \Psi, A \Psi \ra / ||\Psi||^2$ are unaffected by the multiplication of $\Psi$ with an overall 
factor, $\Psi \to c \Psi$. The previous cases may be described as follows:
\begin{enumerate}
	\item The activities $(\alpha_n)$ rescale to a probablity distribution on $\NN$. 
		The expected polymer length is finite.
	\item The activities $(\alpha_n)$ rescale to a probablity distribution on $\NN$
		with infinite expectation value. 
	\item The activities cannot be rescaled to a probability distribution. 
\end{enumerate}
The last two cases are slightly pathological, and we would like to exclude them. This can actually be done, 
for large $\gamma$ (thin cylinders), by a perturbative argument. What happens for small $\gamma$ remains open.

Rezayi and Haldane~\cite{rh94} have observed that for a fixed, finite number of particles, 
the limit $\gamma \rightarrow \infty$ of Laughlin's wave function gives the 
Tao-Thouless state; in block notation, this is a pure trimer state $(010\mid 010\mid... \mid 010)$. 
In the polymer picture, one can show that the activities satisfy
\begin{equation} \label{eq:activasy}
	\alpha_1 = 1,\quad \alpha_n = O\bigl (e^{-4(n-1) \gamma^2}\bigr)\quad (\gamma \to \infty), 
\end{equation}
see also Table~\ref{tab:ux}. Hence, 
in the limit of infinitely thin cylinders ($\gamma \rightarrow \infty$),
 all activities except $\alpha_1$ vanish.
In this limit, $C_N = 1$, $r=1$, and we are in the first case, 
i.e., $q=\lim r^NC_N >0$. One can show that this survives for 
sufficiently large $\gamma$ (see \cite{jls} for details).\\

\textbf{A computational scheme.} Closely related to Eq.~(\ref{eq:activasy}) is the observation that 
long polymers have a small activity. 
This suggests approximating 
Laughlin's state by a quantum polymer with polymers that cannot become longer than $3m$, for some 
cutoff parameter $m \in \NN$. The normalization of such a cutoff polymer system 
satisfies a recurrence relation of finite order, and the normalization can be computed explicitly. 

We are thus let to the following procedure for computing approximations $r_m,\,q_m$ to the quantity $r,\,q$ that 
appears in the asymptotics of $(C_N)$. 
\begin{itemize}
	\item Compute the normalization of Laughlin's function for up to $m$ particles, $C_1,...,C_m$. 
	\item Compute the activities  $\alpha_1,..., \alpha_m$ of polymers of length smaller or equal to $3m$,
		by using the recurrence relation 
		\begin{equation} \label{eq:renewal}
		 	C_n=\alpha_1 C_{n-1}+...+\alpha_{n-1} C_1 + \alpha_n\quad (n\geq 2).
		\end{equation}
	\item Solve the equation  
		$ \alpha_1 t + \alpha_2 t^2 + \alpha_3 t^3+...+ \alpha_m t^m =1$, $t>0$.
		The unique solution $r_m$ is an approximation to $r$. 
		An approximate value to $q = \lim_{N\to \infty} r^N C_N$ is given by 
		$q_m:=\sum_1^m n \alpha_n r_m^n$. 

\end{itemize}
For each fixed $\gamma$, the approximations $r_m, q_m$ converge to $r,q$ when 
the cutoff parameter $m$ goes to infinity. For sufficiently large $\gamma$, we can say 
something about the speed of convergence:
 		$$ r_m - r = O(e^{-4(m-1)\gamma^2}) \quad (m\to \infty)$$
and similarly for $q_m$. 

Roughly, this procedure is a way of computing the expansion of $r$ in powers of 
$\exp(-\gamma^2)$. The interesting point is that each approximate value $r_m$ only requires 
knowledge of Laughlin's wave function for a \emph{finite} number of particles. 
Notice also that in this scheme the Tao-Thouless state, with normalization $1$, corresponds to the 
first approximation, $m=1$ (only trimers allowed). 

In the next section, we will see that something similar can be done for the one-particle density 
and the occupation numbers. 

\section{One-particle density}\label{sec:density}

Now we explain how the polymer representation and the 
results on the normalization lead to symmetry breaking, and how the periodic 
density can be computed. 

Recall that the one-particle density 
$\rho_N(z)$ is obtained from $|\Psi_N(z_1,...,z_N)|^2$ by integrating 
out all variables but one, and normalizing in such a way that $\int \rho_N(z)\intd x \intd y = N$. 
As usual, let $\lfloor N/2 \rfloor$ denote the largest integer below $N/2$. 
In order to eliminate boundary effects, we take the limit of long cylinders
(fixed radius), and obtain a periodic density:
\begin{quote}\it
 	For sufficiently large $\gamma$, the one-particle density $\rho_N(z)$ 
	of Laughlin's wave function~(\ref{eq:laufunct}) has a limit $\rho(z)$ 
	when $N\rightarrow \infty$  for fixed radius $R = 1/\gamma$: 
	\begin{equation*}
	 	\rho(z) = \lim_{N\rightarrow \infty} \rho_N(z-3\lfloor N/2 \rfloor \gamma).
	\end{equation*}
	The limiting density $\rho(z)$ is periodic in the direction along the cylinder axis. The 
	smallest period is $3\gamma$. 
\end{quote}
The shift $3 \lfloor N/2 \rfloor \gamma$
 ensures that  we look at the density near the middle of the cylinder, far away from 
boundaries. -- A formal proof can be found in \cite{jls}. Here, we shall focus 
on the resulting representation of the periodic density $\rho(z)$. 

Let $\hat n_k$ be the number operator for the lowest Landau level 
basis function $\psi_k(z)$. In the limit $N\rightarrow \infty$, the density $\rho(z)$ 
is a sum of Gaussians weighted by the occupation numbers $\la \hat n_k \ra$:
\begin{equation} \label{eq:density}
 	\rho(z) = \sum_{k=-\infty}^\infty \la \hat n_k \ra |\psi_k(z)|^2 
		\propto \sum_{k=-\infty}^\infty \la \hat n_k \ra \exp[-(x-k\gamma)^2].
\end{equation}
A similar formula holds for $\rho_N(z)$ (finite $N$); in this case the summation goes 
from $0$ to $3N-3$.

When $\gamma$ is large enough so that the limit $q$ of $r^N C_N$ is strictly positive, 
there is a nice formula for the infinite cylinder occupation numbers $\la \hat n_k\ra$:
\begin{equation} \label{eq:occnuminf}
	\la \hat n_k \ra = \sum_{X} \rho^\mathrm{P}(X) \la \hat n_k \ra_X, \quad 
			\la \hat n_k \ra_X = \la u_X, \hat n_k u_X\ra/||u_X||^2.
\end{equation}
The sum is over polymers 
\begin{equation*}
	X = \{3j-1,3j,...,3(j+n-1)+1\},\quad j \in \ZZ,\ n\in \NN,
\end{equation*}
and the ``polymer correlation''
\begin{equation} \label{eq:polcorinf}
 	\rho^\mathrm{P}(X) := q r^{N(X)} \alpha_{N(X)}
\end{equation}
is interpreted as the probability of finding the rod $X$. The number $\la \hat n_k \ra_X$ 
may be thought of as the probability of finding a particle in the lattice site 
$k$ given that $k$ is covered by the polymer $X$. 

Eqs.~(\ref{eq:occnuminf}) and~(\ref{eq:polcorinf}) are proved in \cite{jls}. Here, let us 
only mention that $\rho^\mathrm{P}$ is uniquely determined by the following criteria:
\begin{itemize}
	\item It is $3$-periodic: $\rho^\mathrm{P}(X+3) = \rho^\mathrm{P}(X)$. This reflects 
		the covariance~(\ref{eq:covar}).
	\item The probability for finding a polymer $X$ is proportional to the probability 
		$r^{N(X)} \alpha_{N(X)}$ (recall from Sec.~\ref{sec:norm} that $(r^n \alpha_n)$ 
		defines a probability measure on $\NN$). 
	\item The polymer correlation $\rho^\mathrm{P}(X)$ satisfies the sum rule
		 	$\sum_{X\ni 0} \rho^\mathrm{P}(X) = 1.$ The sum rules expresses
		 that each lattice site (in particular, $0$)
		is covered by exactly one polymer 
		\cite{gruberkunz}.
\end{itemize}

Through Eq.~(\ref{eq:occnuminf}), the occupation numbers and the one-particle density 
inherit the $3$-periodicity~(\ref{eq:covar})  from the polymer system: 
\begin{equation*}
	\la \hat n_{k+3} \ra = \la \hat n_k \ra, \quad \rho(z-3\gamma ) = \rho(z).
\end{equation*}
Thus $3$ (resp. $3\gamma$) is \emph{one} period; but is it the \emph{smallest} period? 

The answer is yes. Again, the proof goes with a perturbative argument. Because of
Eq.~(\ref{eq:activasy}), only trimer blocks ``010'' survive in the limit $\gamma \rightarrow \infty$.
 As a consequence, \cite{jls}
\begin{equation} \label{eq:densasy}
		\la \hat n_k \ra = \begin{cases}
				1+ O\bigl(\exp(-\gamma^2)\bigr), & \text{if}\ k \in 3\ZZ,\\
				O\bigl(\exp(-\gamma^2)\bigr), & \text{else}.
			\end{cases}
\end{equation}
This generalizes a similar statement on the occupation numbers for a fixed, 
finite number of particles \cite{rh94}. 
It follows from Eq.~(\ref{eq:densasy}) that
 for large $\gamma$, the smallest period of $\rho(z)$ is  $3\gamma$. 
Numerical results \cite{swk04} show that the period is $3\gamma$ for all 
values of $\gamma>0$, although the amplitude of oscillations becomes very small when $\gamma$ is small. 

\begin{remark}[Average density, reversal invariance]
	With our formulas one can check that the one-particle density oscillates around 
	the average value $(3\cdot 2\pi)^{-1} = (3\cdot 2\pi \ml^2)^{-1}$, which is the correct value for 
	filling factor $1/3$. -- The reversal invariance (see Rem.~\ref{rem:revinv}) leads to a
	a symmetry with respect to the origin: we have
	 $\la\hat n_{-k} \ra = \la \hat n_k\ra$, $\rho(-z)=\rho(z)$ (see also~\cite{swk04}). \\
\end{remark}

\textbf{An approximation scheme for occupation numbers and the one-particle density.}
In Sec.~\ref{sec:norm}, we have explained how we can compute approximations to the quantities 
involved in the asymptotics of the normalization constants by cutting off long polymers. 
Something similar can be done for the occupation numbers $\la \hat n_k \ra$. Notice that, 
because of Eq.~(\ref{eq:density}), computing occupation numbers and the one-particle density is 
essentially the same. 

Let $m\in \NN$ be a cutoff parameter as in Sec.~\ref{sec:norm}. We consider the quantum polymer 
system obtained from Laughlin's state by discarding all polymers with length strictly greater than 
$3m$. In the limit of long cylinders, the one-particle density of this truncated system is periodic, 
regardless of the size of $\gamma$. The
 occupation numbers are obtained by modifying Eqs.~(\ref{eq:occnuminf}) and~(\ref{eq:polcorinf})
 as follows: In Eq.~(\ref{eq:occnuminf}), the summation is restricted to rods $X$ with 
length $|X| \leq 3m$, and in Eq.~(\ref{eq:polcorinf}), $r$ and $q$ are replaced with their 
truncated values $r_m$ and $q_m$ as defined at the end of Sec.~\ref{sec:norm}. 

In this way, for each $m$ and every fixed $\gamma$, we obtain  approximate
occupation numbers $\la \hat n_k \ra ^{(m)}$. In order to compute the $m$'th 
approximation, 
 we need only know Laughlin's wave function for up to $m$ particles. 
Again, the first approximation 
($m=1$) is just the Tao-Thouless state with occupation numbers $1$ at multiples of $3$ 
and $0$ otherwise. Approximations for larger $m$ look more complicated, but for each $m$, 
the approximate occupation numbers are $3$-periodic. 

It remains to see if this approximation scheme converges. For the normalization, 
we know that $r_m \to r$, $q_m\to q$ for each fixed cylinder radius. The answer 
for the occupation numbers depends on the validity of Eq.~(\ref{eq:occnuminf}): so far we only know 
that Eq.~(\ref{eq:occnuminf}) holds when $\gamma$ is large enough. In this case, the approximate occupation 
numbers do converge to the occupation numbers in Laughlin's state, and the speed of convergence is 
\begin{equation*}
 	\la \hat n_k\ra ^{(m)} - \la \hat n_k \ra = O(e^{-4(m-1)\gamma^2}) \quad (m\to \infty), 
\end{equation*}
which complements Eq.~(\ref{eq:densasy}).

\begin{acknowledgements}
 E.H.L. thanks the Alexander von Humboldt Foundation for a research award, based at the
Technical University, Berlin. He also thanks the US National Science
Foundation for partial support, grant PHY-0652854. S.J. and R.S. thank the 
DFG for support under grant no. SE 456/7-1 under the priority program SPP 1092.
\end{acknowledgements}

%% Bibliography %%%%%%%%%%%%%%%%%%%%%%%%%%%%%%%%%%%%%%%%%%%%%%%%%

% \bibliographystyle{unsrt}
% \bibliography{biblio.bib}

\end{document}